# Design and First Results of COFFEE3: A 55nm HVCMOS Pixel Sensor Prototype for High-Energy Physics Applications

**Xiaomin WEI,[a] Zijun XU,[b*] Weiguo LU,[b] Yang ZHOU,[b,1] Zhan SHI,[c] Leyi LI,[b,d] Xiaoxu ZHANG,[b,e] Pengxu LI,[f] Jianpeng DENG,[f] Yang CHEN,[c] Yujie WANG,[c] Zhiyu XIANG,[g] Mei ZHAO,[b] Cheng ZENG,[b,h] Mengke CAI,[b] Boxin WANG,[b,h] Yuman CAI,[b,h] Bingchen YAN,[b,h] Anqi WANG,[h] Yu ZHAO,[a] Zexuan ZHAO,[a] Zheng WEI,[a] Huimin WU,[a] Ruiguang ZHAO,[a] Hongbo ZHU,[f] Yongcai HU,[a] Jianchun WANG[b,i] and Yiming Li[b,i]**

[a] *Northwestern Polytechnical University, Xi'an, China*
[b] *Institute of High Energy Physics, Beijing, China*
[c] *Dalian Minzu University, Dalian, China*
[d] *Shandong University, Qingdao, China*
[e] *Nanjing University, Nanjing, China*
[f] *Zhejiang University, Hangzhou, China*
[g] *Central South University, Changsha, China*
[h] *University of Chinese Academy of Sciences, Beijing, China*
[i] *High Energy Physics Research Center of Henan Academy of Sciences, China*
*E-mail:* zhouyang@ihep.ac.cn

ABSTRACT: Motivated by the stringent requirements of the Upstream Pixel (UP) tracker in the LHCb Upgrade II and the Inner Tracking detector (ITK) of the Circular Electron Positron Collider, the COFFEE series of pixel sensor chips have been developed using a 55nm High-Voltage CMOS (HVCMOS) process. The primary objective is to achieve a time resolution of a few nanoseconds under a hit density of up to 100 MHz/cm$^2$, while maintaining fine spatial resolution (~10 μm) and reasonable power consumption (<200 mW/cm$^2$). Building on the process validation of the COFFEE2 prototype, this work presents the design and preliminary test results of COFFEE3—a prototype integrating two distinct readout architectures. Architecture 1, tailored for the current triple-well process, adopts NMOS-only in-pixel circuitry and innovative column-level readout to handle high hit densities. The time walk of pixel-level signal is controlled within 10 ns, and the Time of Arrival (TOA) and Time over Threshold (TOT) are measured with a system clock with the period of 25 ns in peripheral circuits. Architecture 2, developed for future possible processes with p-type buried layer isolation, features pixel-level time measurement and storage. A chip-level Time-to-Digital Converter (TDC) is used and the part of Voltage-Controlled Delay Line (VCDL) is copied in each pixel to get a high time resolution. The TOA resolution is estimated to be 4.2 ns and the TOT resolution 8.4 ns. COFFEE3, with a layout size of 3×4 mm$^2$, was manufactured and has undergone preliminary tests. Charge injection tests for analog circuits, and laser tests for full readout chains, confirm that both architectures operate as expected. Next step work will focus on characterizing key performance such as the timing resolution, radiation hardness, and tracking performance of minimum ionising particles.

---

[*] Equal first authors.
[1] Corresponding author.

KEYWORDS: HVCMOS; pixel sensors; TDC; high hit density; time measurement.

## Contents

## 1. Introduction

High-energy physics (HEP) experiments, such as the LHCb Upgrade II and planned future Circular Electron-Positron Colliders (CEPC), demand advanced pixel tracking systems to capture the trajectory of charged particles with unprecedented precision. As for the Upstream Pixel (UP) tracker, the pixel tracking systems in the LHCb Upgrade II face challenges imposed by high-luminosity operation conditions: radiation hardness to withstand a neutron fluence of up to $3\times10^{15}$ $n_{eq}/cm^2$; capability to resolve extreme hit density of 100 MHz/cm²; a few second time resolution to distinguish the neighboring bunch crossing of 25 ns at the LHC; fine spatial resolution of about 10 μm in the direction to accurately reconstruct particle trajectories that bends in the magnetic field [1]. In addition to a similar time resolution on the order of ~ns, the CEPC inner tracker (ITK) has higher requirements for position resolution, with a unidirectional position < 10 μm [2].

The feasibility of HVCMOS pixel sensors for high radiation tolerance has been validated by several chips, such as ATLASPix [3] and MuPix [4]. Other sensors are being developed for future applications, such as MightyPix series [5-6] and RD50-MPW series [7]. MightyPix is derived from ATLASPix for the LHCb downstream tracker upgrade, which aims for a time resolution of 3 ns using a 40 MHz timestamp clock [8]. Comparing to the 180 nm or 150 nm processes used for the sensor chips mentioned, more advanced 55 nm HVCMOS process would allow the integration of complex analog and digital circuits within a compact area, addressing the need to provide high-precision time and position resolution in high hit density applications. COFFEE2, the first validation prototype using 55nm HVCMOS process, is designed and produced for exploring and verifying the characteristics of the process, including the performance of passive sensing diodes and in-pixel amplification circuits [9]. Building on the success of COFFEE2, COFFEE3 is designed to address two key goals, which are: (1) to verify different readout circuit structures compatible with either the current triple well or future possible processes with a p-type isolation between the readout circuits and the charge collection node, and (2) to characterize the time resolution, power consumption, radiation hardness, and tracking performance of minimum ionising particles.

The rest of this paper is organized as follows. Section 2 details the architecture design of COFFEE3, particularly the in-pixel time measurement architecture. Section 3 presents preliminary test results. Section 4 summaries the work.

## 2. COFFEE3 Design Details

### 2.1 Overall Design

The design of COFFEE3 aims at the common requirements of the LHCb UP upgrade and the CEPC ITK. The pixel layout height is set to 40 μm, which leads to an actual pixel size of 36 μm (the manufacturing size of the 55nm process will be shrunk to 0.9 times relative to the layout size). Without considering charge sharing, the expected position resolution achieved in this direction is ~10 μm (36/√12). The time resolution is targeted at <5 ns to ensure 99% tagging accuracy of the 25 ns bunch spacing at the LHC. To keep power dissipation below 200 mW/cm² for the entire chip, the power consumption budgets for the pixel array and the peripheral circuits are approximately 160 mW/cm² (~140 mW/cm² for analog parts and ~20 mW/cm² for digital parts) and 40 mW/cm², respectively.

COFFEE3 includes two different pixel array readout structures and their matching peripheral digital circuits. The two readout architectures are designed respectively for the current triple-well process (as shown in Figure 1(a)) and the desired process with a p-type buried layer isolation (as shown in Figure 1(b)), for which process modification is being pursued.

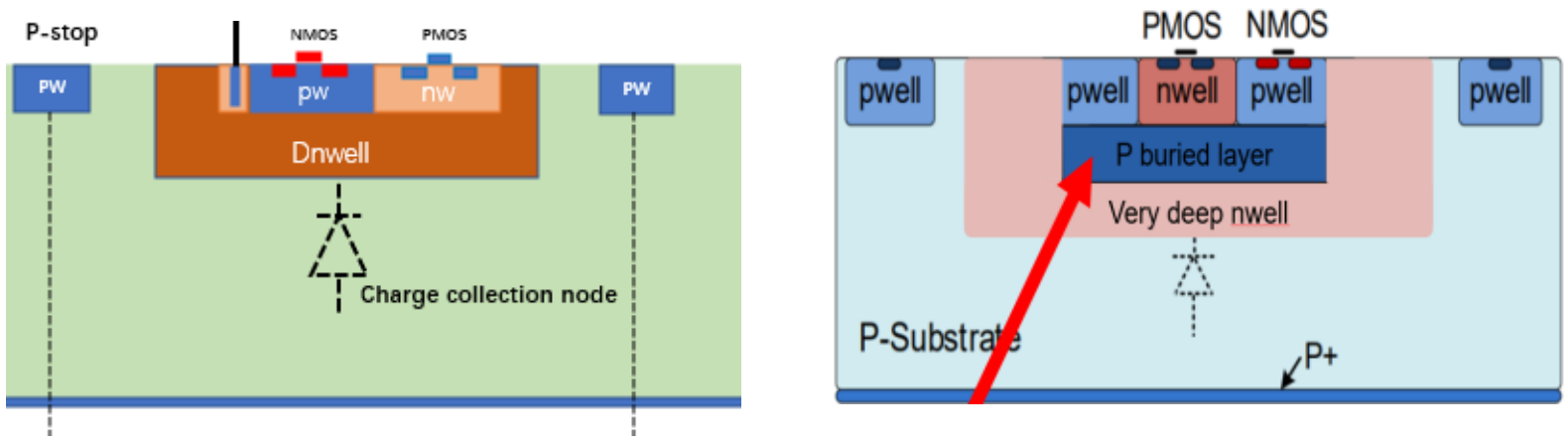

(a) (b)

**Figure 1.** Process cross-section view of (a) the current triple well process and (b) future possible process with p-type buried layer isolation

Architecture 1 is optimized for the current triple-well 55nm HVCMOS process. In triple-well processes, deep-Nwell sensing diodes are prone to crosstalk with Nwell-hosted PMOS transistors; thus, an NMOS-only in-pixel design is adopted to mitigate this issue. This architecture draws inspiration from existing prototypes like ATLASPix [3] and MightyPix [5-6], but introduces modifications to enhance hit density handling. The design is expected to result in a small circuit area, which means small diode capacitance, thereby enabling small time-walk and a good signal-to-noise ratio with adequate power consumption.

Architecture 2 is developed for the desired process, which introduce P-type buried layer between the Nwell and the sensing diode, eliminating crosstalk between PMOS and sensing diodes. This enables a CMOS in-pixel design, supporting more complex circuitry. Specifically, pixel-level fine time quantization and storage can be implemented. As a result, high hit density and time resolution can be achieved.

Figure 2 presents the block diagrams of the two readout architectures. The following sections illustrate the details.

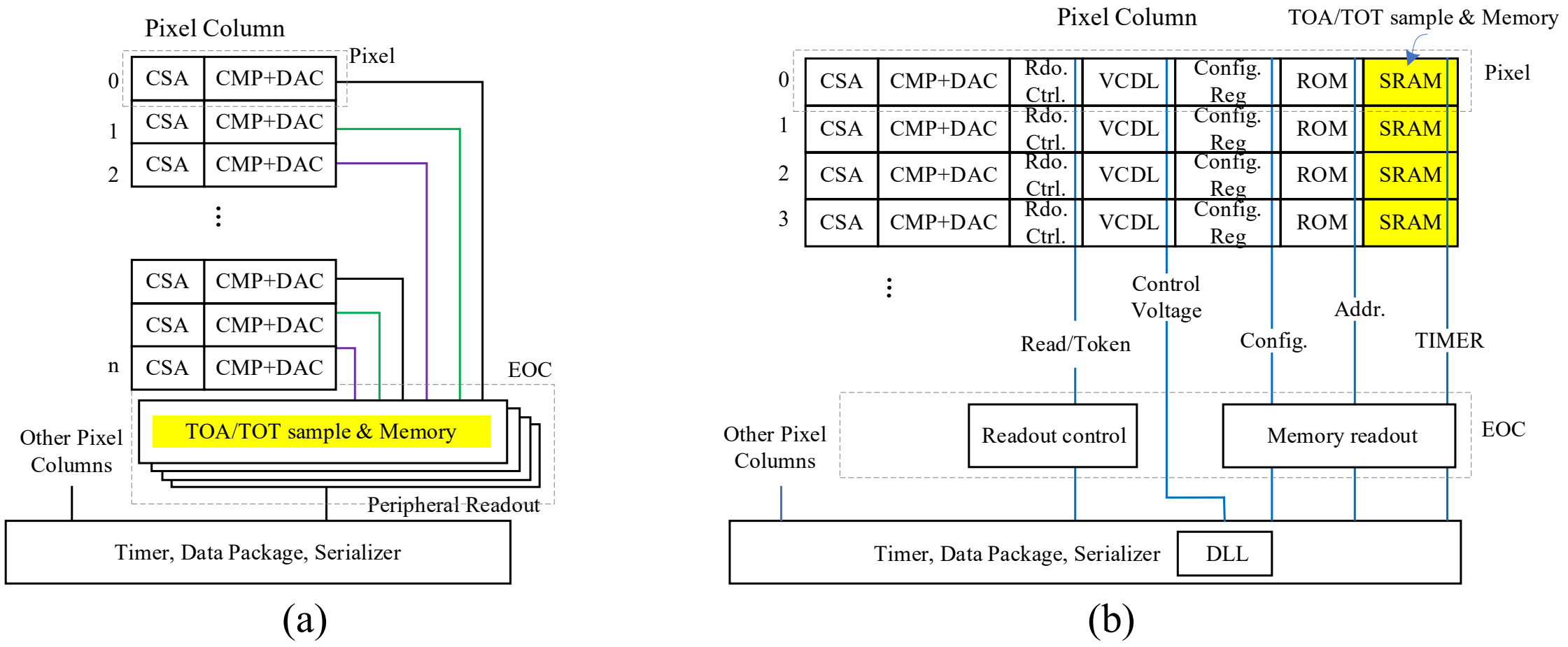

**Figure 2.** Block diagrams of (a) the architecture 1 for triple well process and (b) the architecture 2 for the process with p-type buried layer isolation

### 2.2 Architecture 1 with NMOS-only in-pixel circuitry

Figure 2(a) presents the architecture for the current triple-well process. The pixel integrates a charge-sensitive amplifier (CSA), a comparator (CMP), and a 4-bit Digital Analog convertor (DAC) to mitigate threshold mismatch from pixel to pixel. All comparator outputs are routed to End-of-Column (EOC) modules. A single column typically contains hundreds of pixel outputs for full size sensor chips. If the hit density is low, the column pixels are rarely hit frequently in a short period. As a result, these outputs first undergo an OR operation, and the leading and trailing edge times of the resulting hit signal are sampled and processed. Time of Arrival (TOA) is derived from the leading edge, and the Time over Threshold (TOT) is the difference between leading edge and trailing edge. Generally, TOT helps correction of TOA error induced by time walk.

To meet the hit density requirement of LHCb UP upgrade II and address sequential same-column hits (critical for high-event-rate handling), all the columns are read out in parallel and two complementary optimizations are adopted. Firstly, the pixels in one column are split into several groups for parallel readout, and grouping is optimized for simultaneous hits (e.g., 4 adjacent pixels assigned to 4 groups) to maximize throughput and to reduce pile-up. Secondly, each group uses a two-stage pipelined signal processing unit, which is able to record and store the next hit signal while transmitting the current one, cutting dead time in signal processing and further minimizing pile-up.

### 2.3 Architecture 2 with pixel-level time measurement

Taking the advantages of the 55nm process, Architecture 2 implements coarse and fine time quantization within the pixel. Figure 2(b) shows the block diagram. The pixel structure includes an analog section with a CSA, a comparator, and an in-pixel DAC, as well as a digital section comprising a readout controller, a Voltage-Controlled Delay Line (VCDL), pixel configuration circuits, pixel address ROM cells, and SRAM cells for sampling and storage of time information. The hit information is first stored in the pixel and then read out to the matrix periphery in order of priority for each double column. Memory access circuits at the EOC are required for all types of memories. A data-driven readout circuit is implemented in our work, based on the principle proposed in Reference [10].

Coarse time quantization is generated by counting the 40 MHz system clock (CLK). The pixel samples and stores this count (TIMER) at the leading and trailing edges of the comparator output signal (DIS_OUT), respectively. SRAM is used for the timestamp storage (sampled from TIMER lines), taking into account the power consumption associated with signal transitions. Figure 3 illustrates the differences between distributing a single TIMER line (TS) to an SRAM cell and to a receiving buffer. The transition time of TS is large due to the significant parasitic load on TS in a larger pixel array. When TS is received by a buffer, the transition current ($I_{trans}$) is huge from Mb1 to Mb2. In contrast, only leakage current ($I_{leak}$) exists on Ma1 when connected to an SRAM cell.

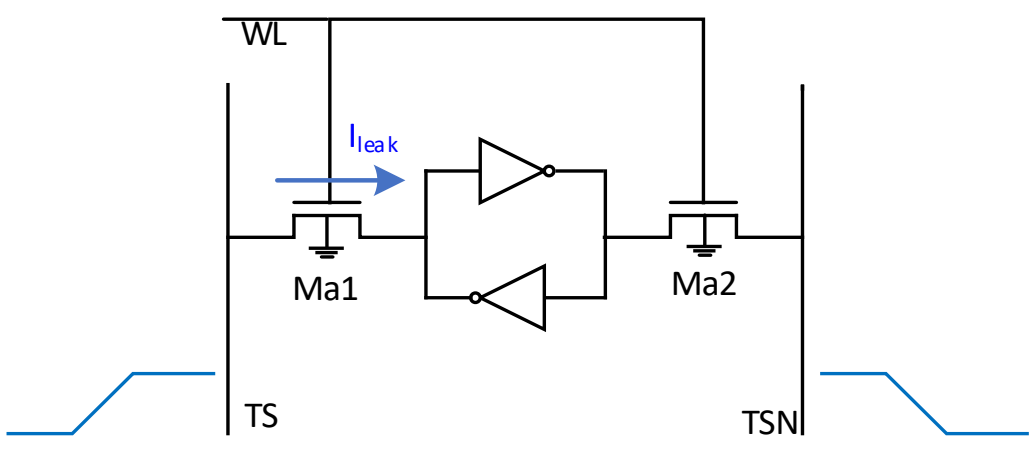

(a) (b)

**Figure 3.** Distribution of a single TIMER line (TS) (a) to an SRAM cell and (b) to a receiving buffer

The fine time quantization is implemented via a two-part Time-to-Digital Converter (TDC). Figure 4 presents the schema and the timing sequence for the fine time measurement scheme. A Delay-Locked Loop (DLL) is deployed in the peripheral circuitry of the pixel array; once locked, its VCDL uniformly divides one clock cycle into six phases. A replica of the DLL's VCDL is integrated into each pixel, and using the control voltage from the locked DLL, the in-pixel VCDL also divides one clock cycle into six equal parts. To reduce power consumption, the DIS_OUT signal but not the clock signal passes through the VCDL to generate phase shifts, and the phase-shifted data is sampled at the rising edge of the next clock cycle to obtain the time difference between DIS_OUT and the clock [11]. In this design, five phases are sampled for the leading edge of DIS_OUT achieving a time precision of 25/6 ns (~4.2 ns). For the trailing edge, 2 phases are sampled, resulting in a time precision of 25/3 ns (~8.4 ns).

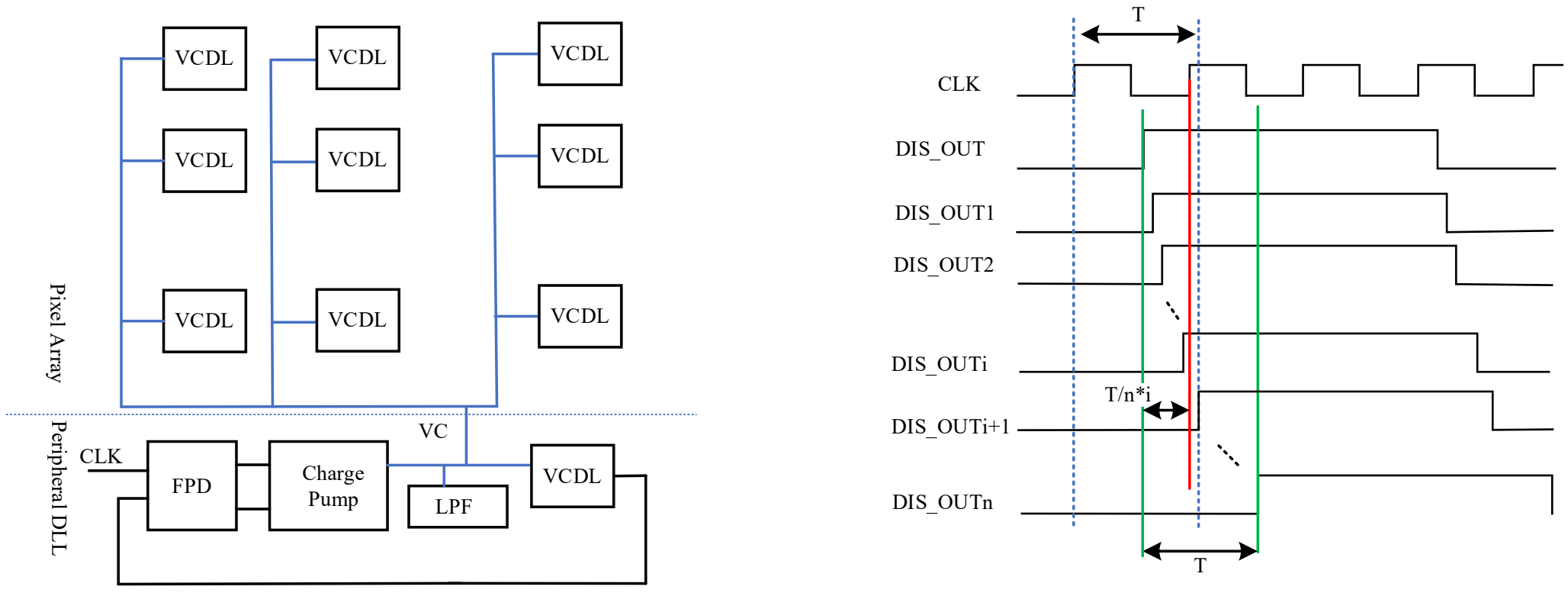

(a) (b)

**Figure 4.** Schema and the timing sequence of the fine time measurement scheme

## 3. Preliminary Test Results

COFFEE3 was submitted for fabrication in January 2025 and returned in May 2025. The data acquisition system mainly includes a COFFEE3 chip carrier board, an AMD ZCU102 evaluation board, and an FMC mezzanine card which is named CaR board v1.5 from Caribou DAQ system Project [12].

The analog pixel circuits (CSA and Comparator) were tested with charge injection by signal generator. The DLL delivers the clock phase delay as expected. Each tap shifts by a phase of $\pi/3$. The full readout chains were validated by laser test. Figure 5 presents a result of charge injection test for Architecture 1. The data format and the address information are correct. Figure 6 presents a laser test of Architecture 2. TOT values were measured from comparator outputs with different thresholds. The results match the circuit simulation. The valid transmission packet contains the correct row and column addresses corresponding to a particle hit. The full readout chain was validated.

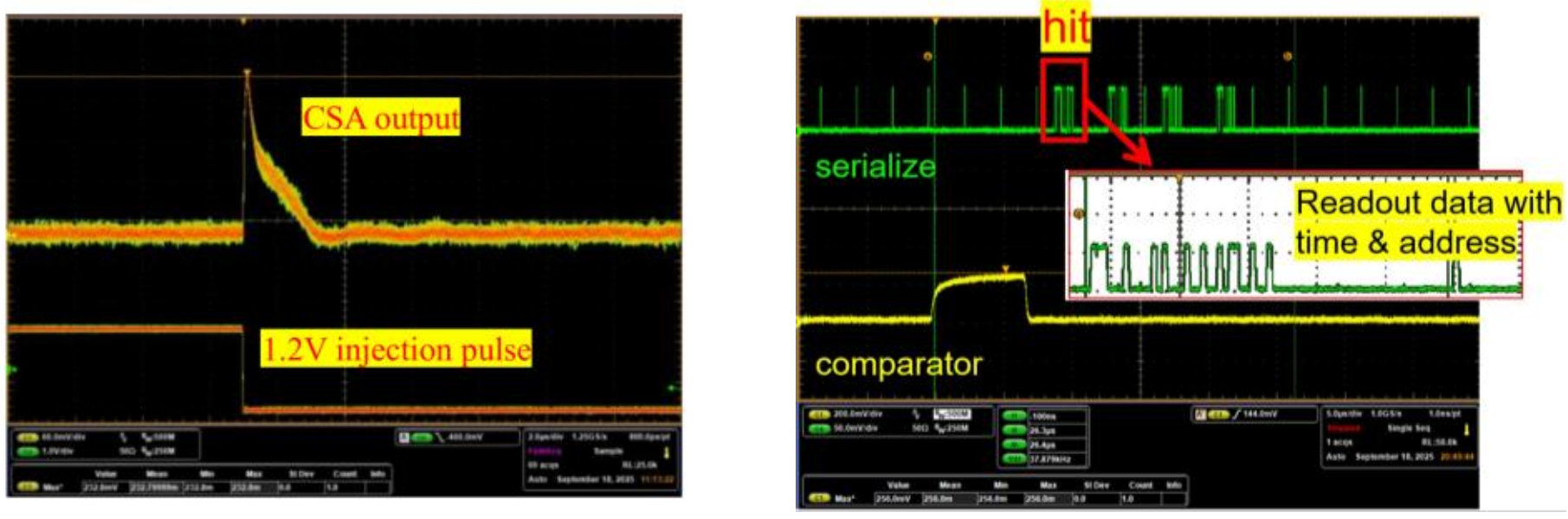

(a) (b)

**Figure 5.** Charge injection test results of Architecture 1. (a) Injection pulse and CSA output; (b) Serial data output

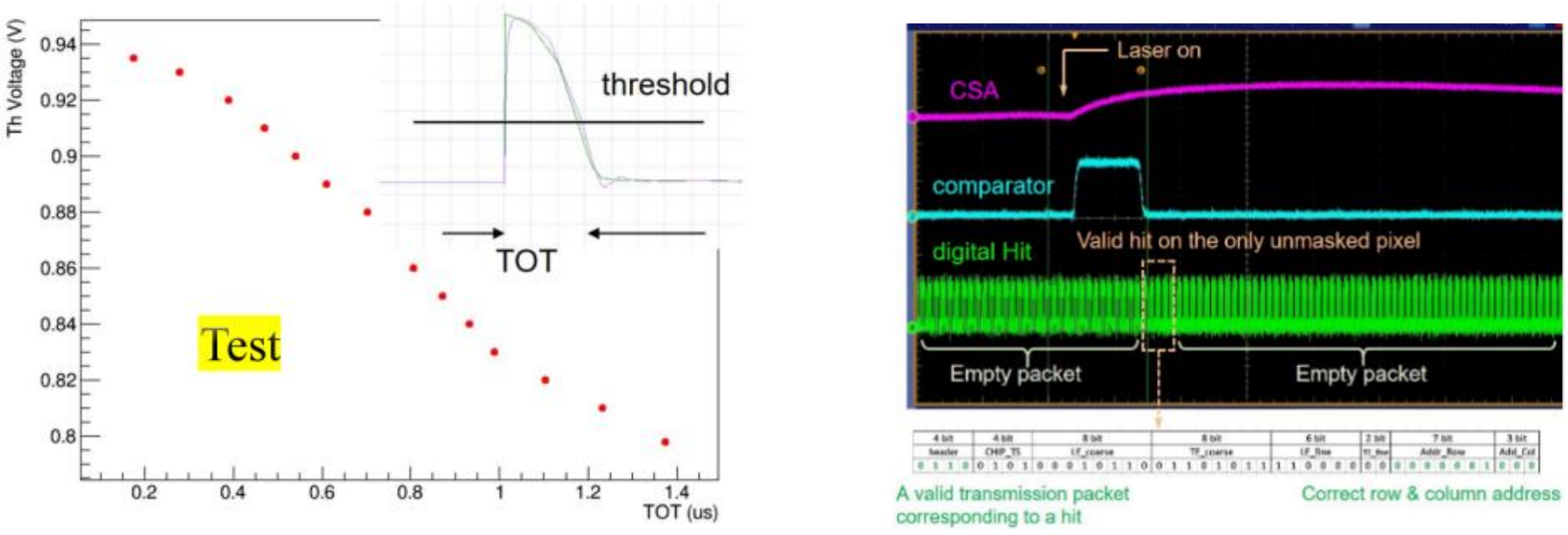

(a) (b)

**Figure 6**. Laser test results of Architecture 2. (a) TOTs measured from a comparator; (b) A valid transmission packet containing the correct row and column addresses corresponding to a hit.

Further characterization of the chip performance is ongoing, such as details of TOA and TOT with X-ray from $^{55}$Fe source, beta-ray (electron) from $^{90}$Sr source, and minimum ionizing particles from cosmic-ray or accelerator. The irradiation and beam tests are planned. The time resolution, the spatial resolution, hit efficiency, and charge collection efficiency will be measured in the future.

## 4. Conclusion

This work presents the design and preliminary test results of COFFEE3, a 55nm HVCMOS pixel sensor prototype developed for high-energy physics applications. Two distinct architectures are integrated to adapt to current and future possible HVCMOS processes. Architecture 1 (NMOS-only in-pixel design) is optimized for the current triple-well process, with modifications to enhance hit density handling through column-level pixel grouping and pipelined two-stage signal processing. Architecture 2 (pixel-level time measurement) is tailored for future possible modified process with p-type buried layer isolation, implementing in-pixel TDC functionality to achieve an LSB of ~4.2 ns.

Preliminary tests confirm that both architectures operate as expected. Future work will focus on comprehensive performance characterization, including time resolution, detection efficiency, radiation hardness, and high energy test beam validation, serving as a solid step for the LHCb Upgrade II UP tracker and the CEPC ITK developments.

## Acknowledgments

This work is supported by National Natural Science Foundation of China under Grant No. W2443008, No. 12375191, No. 12435013, No. 12475195; the National Key Research and Development Program of China under Grant No. 2023YFA1606300, No.2023YFE0206300, No. 2023YFF0719600, No. 2024YFE0110102; Guangdong Basic and Applied Basic Research Foundation under Grant NO. 2024A1515012141; China Postdoctoral Science Foundation under Grant No. 2023M742850.